\documentclass[twocolumn,showpacs,prb,amsfonts,amsmath,amssymb,floatfix]{revtex4}  

\usepackage{color}
\usepackage{hhline}
\usepackage{mathrsfs}
\usepackage{graphicx}
\usepackage{dcolumn}
\usepackage{bm}
\usepackage{multirow}
\usepackage{booktabs}
\input{epsf}

\def\C60{C$_{60}$}
\def\K3C60{K$_{3}$C$_{60}$}
\def\Rb3C60{Rb$_{3}$C$_{60}$}
\def\Cs3C60{Cs$_{3}$C$_{60}$}
\def\A3C60{A$_{3}$C$_{60}$}
\def\pic1{picene$^{3-}$}
\def\cor1{coronene$^{3-}$}
\def\phe1{phenanthrene$^{3-}$}

\def\t1u{$t_{{\rm 1u}}$}
\def\Vopt1{$V_{\rm{SC}}^{{\rm{opt.}}\ \!\!P}$}
\def\Vsmall1{$V_{\rm{SC}}^{{\rm{high}}\ \!\!P}$}
\def\VMIT1{$V_{\rm{MIT}}$}
\def\VAFI1{$V_{\rm{AFI}}$}
\begin{document}

\title{{\em Ab initio} derivation of electronic low-energy models for C$_{60}$ and aromatic compounds}

\author{Yusuke Nomura$^{1}$}
\author{Kazuma Nakamura$^{1,2}$}
\author{Ryotaro Arita$^{1,2,3}$}
\affiliation{$^1$Department of Applied Physics, University of Tokyo, 7-3-1 Hongo, Bunkyo-ku, Tokyo, 113-8656, Japan} 
\affiliation{$^2$JST CREST, 7-3-1 Hongo, Bunkyo-ku, Tokyo, 113-8656, Japan}
\affiliation{$^3$JST PRESTO, Kawaguchi, Saitama, 332-0012, Japan} 

\date{\today}

\begin{abstract}
We present a systematic study for understanding the relation between electronic correlation and superconductivity in C$_{60}$ and aromatic compounds. We derived, from first principles, extended Hubbard models for twelve compounds; fcc \K3C60, \Rb3C60, \Cs3C60 (with three different lattice constants), A15 \Cs3C60 (with four different lattice constants), doped solid picene, coronene, and phenanthrene. We show that these compounds are strongly correlated and have a similar energy scale of the bandwidth and interaction parameters. However, they have a different trend in the relation between the strength of electronic correlation and superconducting transition temperature; while the C$_{60}$ compounds have a positive correlation, the aromatic compounds exhibit negative correlation.
\end{abstract} 
\maketitle 

\section{Introduction} \label{sec:intro}
Superconductivity in  $\pi$-electron systems, whose history dates back to studies on graphite intercalation compounds in 60's,\cite{1st} has attracted broad interest in condensed matter physics. Recently, two seminal discoveries for carbon-based superconductors have been reported. One is A15/fcc \Cs3C60, with a new method to synthesize highly crystalline samples.\cite{A15Cs1,A15Cs2,fccCsN,Cs-NMR} The observed superconducting transition temperatures ($T_c$) for the high-pressure samples are found to be as high as 38 K for A15 and 35 K for fcc. 
The other is potassium-doped solid picene,\cite{pic-Kubozono} which opened a new avenue to various `{\em aromatic superconductors}', for which the maximum $T_c$ has reached 33 K.\cite{aroma-per,phe,phe2,dibenz}

The mechanism of superconductivity of these new superconductors has not been fully understood. For alkali-doped fullerides, while there are several experimental reports which seemingly support the conventional BCS mechanism, various indications for unconventional superconductivity have been also observed. For example, although the positive correlation between $T_c$ and the lattice constant found in K- and Rb-doped fullerides has been understood in terms of the standard BCS theory,\cite{rev-C60} more recent experiments for larger cations have revealed that $T_c$ does not necessarily behave monotonically as a function of the lattice parameters.\cite{A15Cs1,A15Cs2,fccCsN,Cs-NMR} The fact that the superconducting phase has a dome-like shape in the phase diagram is indeed reminiscent of cuprates\cite{cuprates} and unconventional organic superconductors such as BEDT-TTF.\cite{BEDT} 
In addition, these new \C60 superconductors are insulators at ambient pressure,\cite{A15Cs1,A15Cs2,fccCsN,Cs-NMR} indicating that the superconducting phase resides in the vicinity of an insulating phase. In fact, considering these characteristic features, it has been proposed that interplay between orbital, spin, and lattice degrees of freedom is the origin of the high $T_c$ superconductivity.\cite{Capone}

For aromatic superconductors, following the discovery of K-doped picene (C$_{22}$H$_{14}$),\cite{pic-Kubozono} it has been found that various hydrocarbon compounds such as coronene (C$_{24}$H$_{12}$), phenanthrene (C$_{14}$H$_{10}$), and 1,2:8,9-dibenzopentacene (C$_{30}$H$_{18}$) also exhibit superconductivity.\cite{aroma-per,phe,phe2,dibenz} 
Diversity of hydrocarbon molecules 
suggests the possibility of new and higher $T_c$ aromatic superconductors. 
So far, electronic structure,\cite{pic-Kosugi,pic-Kosugi2,pic-band,phe-band,cor-band} electronic correlations,\cite{pic-Capone,pic-Kim} electron-phonon interactions,\cite{pic-Boeri,pic-Mauri} and exciton/plasmon properties of the normal state\cite{pic-Knupfer,pic-Knupfer1,pic-Knupfer2,pic-pla} have been studied, while studies on superconducting properties are still limited. 
The pairing mechanism is totally an open question. 

There are several similarities between \C60 and aromatic superconductors; they are both molecular solids having narrow bands around the Fermi level, whose energy scale competes with that of electron-phonon and electron-electron interactions. 
The competition among these three factors is a characteristic aspect of carbon-based materials. {\em Ab initio} derivations of effective low-energy models for these compounds are important to make the situation transparent and to clarify the origin of their high $T_c$ superconductivity. 
By comparing 
 the parameters in the effective models for the \C60 and aromatic superconductors, 
the differences and similarities 
are quantitatively identified and analyzed.

Recent methodology for construction of the electronic model, based on the combination use of the maximally localized Wannier orbital (MLWO) (Ref.~\onlinecite{maxloc}) and the constrained random phase approximation (cRPA),~\cite{cRPA} extends its applicability range. It does not place limitations on the character of basis orbitals of the effective model, whether atomic or molecular.
Indeed, it has already been applied to various complex systems such as BEDT-TTF (Refs.~\onlinecite{cRPA-ex4} and \onlinecite{cRPA-bedt}) or zeolites.\cite{cRPA-ex5} While electronic interaction parameters of the \C60 and aromatic superconductors have been estimated by various methods,\cite{U1,U2,U3,U4,pic-Knupfer1} explicit and direct comparison of these systems by the same method has yet to be done. Thus it is imperative to evaluate the interaction parameters of \C60 and aromatic superconductors by exploiting the state-of-the-art technique and perform a systematic comparison. 
In the present study, we constructed {\em ab initio} extended Hubbard models which describe the low-energy electronic structure of twelve 
examples of C$_{60}$ and aromatic compounds. The transfer integrals were given as matrix elements of the Kohn-Sham Hamiltonian in the Wannier basis. The interaction parameters were evaluated by calculating the Wannier matrix elements of the screened Coulomb interaction, which is obtained by cRPA.
The estimated correlation strength as the ratio of the interaction energy to the kinetic one is nearly or beyond unity for the studied materials, indicating that both the C$_{60}$ and aromatic systems are classified into a strongly correlated electron systems. On the other hand, we observed a notable difference between the two systems; for the C$_{60}$ system, there exist positive correlation regime in the correlation strength and the experimental $T_{c}$. In contrast, the aromatic system exhibits negative correlation between these two quantities. 

This paper is organized as follows. In Sec.\ref{sec:method}, we show how to construct the low-energy models from {\em ab initio} calculations. 
In Sec.\ref{sec:result}, we show the calculated band structure, 
MLWOs, transfer integrals, and effective interaction parameters. We discuss the material dependence of the derived parameters and relation between the strength of electronic correlation and superconductivity in Sec.\ref{sec:discuss}. Finally we give a summary in Sec.\ref{sec:summary}. 

\section{Methods} \label{sec:method}
We derive electronic low-energy models with the combination of MLWO and cRPA. This method has widely been applied to the derivation of effective models for 3$d$ transition metals,~\cite{cRPA-ex2,cRPA-ex3} their oxides,~\cite{cRPA-ex2} organic conductors,~\cite{cRPA-ex4,cRPA-bedt} zeolites,~\cite{cRPA-ex5} iron-based superconductors,~\cite{cRPA-ex6,cRPA-ex7} and 5$d$ transition metal oxides.\cite{Ir} We first perform band calculations based on density functional theory (DFT),~\cite{DFT1,DFT2} and choose `target bands' of the effective model. By constructing MLWOs for the target bands, we calculate transfer integrals and effective interactions in the effective model. In the calculation of the effective interaction, the screening by electrons besides target-band electrons is considered within cRPA (see below).

We apply this scheme to the derivation of effective models, i.e., extended multi-orbital Hubbard models, for the C$_{60}$ and aromatic compounds. The Hamiltonian consists of the transfer part ${\cal H}_t$, the Coulomb repulsion part ${\cal H}_U$, and the exchange interactions and pair hopping part ${\cal H}_J$ defined as
\begin{eqnarray}
\label{eq1}
{\cal{H}} = {\cal{H}}_t + {\cal{H}}_U + {\cal{H}}_J, 
\end{eqnarray}
where
\begin{eqnarray}
\label{eq2}
{\cal H}_t &=& \sum_{\sigma}\sum_{ij}\sum_{nm} t_{n\!m} \! 
({\mathbf R}_{ij}) a_{i\!n}^{\sigma \dagger} a_{j\!m}^{\sigma}, \\ 
\label{eq3}
{\cal H}_U &=& \frac{1}{2} \sum_{\sigma\rho}\sum_{ij}\sum_{nm}  U_{n\!m} \!
({\mathbf R}_{ij}) a_{i\!n}^{\sigma \dagger} a_{j\!m}^{\rho\dagger} a_{j\!m}^{\rho} a_{i\!n}^{\sigma}, \\ 
\label{eq4}
{\cal H}_J &=& \frac{1}{2}\!\sum_{\sigma\rho}\!\sum_{ij}\!\sum_{nm}\!J\!_{n\!m} \!(\!{\mathbf R}_{i\!j}\!)\!\bigl(\!a_{i\!n}^{\sigma \dagger}\!a_{j\!m}^{\rho\dagger}\!a_{i\!n}^{\rho}\!a_{j\!m}^{\sigma}\!+\!a_{i\!n}^{\sigma\!\dagger}\!a_{i\!n}^{\rho\dagger}\!a_{j\!m}^{\rho}\!a_{j\!m}^{\sigma}\!\bigr)
\end{eqnarray}
with $a_{in}^{\sigma\dagger}$ ($a_{in}^{\sigma}$) being a creation (annihilation) operator of an electron with spin $\sigma$ in the $n$-th MLWO localized at a C$_{60}$ or aromatic hydrocarbon molecule located at ${\mathbf R}_{i}$ and ${\mathbf R}_{ij}$=${\mathbf R}_{i}$$-$${\mathbf R}_{j}$. The parameters $t_{nm}({\mathbf R}_{ij})$ represent an onsite energy (${\mathbf R}_{ij}$=${\mathbf 0}$) and hopping integrals (${\mathbf R}_{ij}$$\neq$${\mathbf 0}$), which are described with the translational symmetry as
\begin{eqnarray}
\label{eq5}
t_{nm}\!({\mathbf R})= \bigl< \phi_{n{\mathbf R}} \bigl | {\cal H}_{KS} \bigr | \phi_{m{\mathbf 0}} \bigr>, 
\end{eqnarray}
where $\bigl| \phi_{n{\mathbf R}_{i}}\bigr>$=$a_{in}^{\dagger}\bigl|0\bigr>$ and ${\cal H}_{KS}$ is the Kohn-Sham Hamiltonian.

To evaluate effective interaction parameters $U_{n\!m}\!({\mathbf R})$ and $J_{n\!m}\!({\mathbf R})$, we calculate the screened Coulomb interaction $W({\mathbf r},{\mathbf r^\prime})$ at the low-frequency limit. We first calculate non-interacting polarization function $\chi$ with excluding polarization processes within the target bands. Note that screening by the target electrons is considered when we solve the effective models, so that we have to avoid double counting of it when we derive the effective models. With the resulting $\chi$, the $W$ interaction is calculated as $W$=$\bigl(1-v\chi\bigr)^{-1}v$, where $v$ is the bare Coulomb interaction $v({\mathbf r},{\mathbf r}^\prime)$=$\frac{1}{|{\mathbf r}-{\mathbf r}^\prime|}$. 

Once screened Coulomb interaction $W({\mathbf r},{\mathbf r}^\prime)$ is calculated, the matrix elements of $W$ are obtained as  
\begin{eqnarray}
U\!_{n\!m} \! ({\mathbf R}) &=& \bigl< 
\phi_{n{\mathbf R}} \phi_{m{\mathbf 0}} 
\bigl | W \bigr| \phi_{n{\mathbf R}} \phi_{m{\mathbf 0}} \bigr > \nonumber \\ 
&=& \int \!\!\! \int \!\!\! d{\mathbf r} d{\mathbf r}^\prime \!
\phi_{n\!{\mathbf R}}^{\ast} (\!{\mathbf r}\!) \phi_{n\!{\mathbf R}} (\!{\mathbf r}\!) W({\mathbf r}\!,\!{\mathbf r}^\prime) \phi_{m\!{\mathbf 0}}^{\ast} (\!{\mathbf r}^\prime\!) \phi_{m\!{\mathbf 0}} (\!{\mathbf r}^\prime\!)
\label{eq6}
\end{eqnarray}
and
\begin{eqnarray}
J\!_{n\!m} \! ({\mathbf R}) &=& \bigl< 
\phi_{n\!{\mathbf R}} \phi_{m\!{\mathbf 0}} 
\bigl | W \bigr| 
\phi_{m\!{\mathbf 0}} \phi_{n\!{\mathbf R}} \bigr > \nonumber \\ 
&=&  \int \!\!\! \int \!\!\! d{\mathbf r} d{\mathbf r}^\prime \!
\phi_{n\!{\mathbf R}}^{\ast} (\!{\mathbf r}\!) \phi_{m\!{\mathbf 0}} (\!{\mathbf r}\!) W(\!{\mathbf r}\!,\!{\mathbf r}^\prime\!) \phi_{m\!{\mathbf 0}}^{\ast}\!(\!{\mathbf r}^\prime\!) \phi_{n\!{\mathbf R}} (\!{\mathbf r}^\prime\!).
\label{eq7}
\end{eqnarray}
For comparison with the cRPA results, we calculate interaction parameters with different levels of screening. One is the unscreened one, i.e., the bare Coulomb interaction, and the other is the fully-screened one where we calculate $\chi$ including the target-band screening. To distinguish these from cRPA, we denote them as `bare' and `fRPA'.

\section{Results} \label{sec:result}
\subsection{Calculation conditions}
\begin{table} 
\caption{
Basic property of fcc and A15 alkali-doped C$_{60}$ compounds; the lattice parameter $a$, corresponding C$_{60}\!^{3-}$ volume in solid, and measured superconducting transition temperature $T_c$ or the N\'{e}el temperature $T_N$. For fcc \Cs3C60, the three samples are specified with the C$_{60}^{3-}$ volume and corresponds to those in the superconducting phase with maximum $T_c$ (\Vopt1), in the vicinity of the metal-insulator transition (\VMIT1), and in the anti-ferromagnetic insulating phase (\VAFI1), respectively. 
For A15 structure, 
we also list another sample with a higher pressure, for which $T_c$ is lowered than that of \Vopt1, and is abbreviated to \Vsmall1.}
\vspace{0.1cm}
\begin{center}
\begin{tabular}{cc@{\  }  c@{\  } c @{\  }c@{\  } c c}
\hline
 &                &  $a$    & Volume/C$_{60}\!\!^{3-}$ &  Pressure   &  $T_c$($T_N$)  &  \multirow{2}{0.45cm}{Ref.} \\
 &                &  (\AA)  & (\AA$^{3}$)        &  (kbar) &   (K)          &   \\
\hline
\multirow{5}{1cm}{\ \  fcc  \A3C60}
        &  K         &  14.240  &    722            &  0      &   19           & \onlinecite{C60-st}      \\
        &  Rb        &  14.420  &    750            &  0      &   29           & \onlinecite{C60-st}      \\
        & Cs(\Vopt1) &  14.500  &    762            &  7      &   35           & \onlinecite{fccCsN}      \\
        & Cs(\VMIT1) &  14.640  &    784            &  2      &   26          & \onlinecite{fccCsN}      \\
        & Cs(\VAFI1) &  14.762  &    804            &  0      &   (2.2)        & \onlinecite{fccCsN}      \\ 
\hline
\multirow{4}{1cm}{\ \ A15 \Cs3C60}
        & \Vsmall1   &  11.450  &    751            &  15     &   35           & \onlinecite{A15Cs2}      \\
        & \Vopt1     &  11.570  &    774            &  7      &   38           & \onlinecite{A15Cs2}      \\
        & \VMIT1     &  11.650  &    791            &  3      &   32       & \onlinecite{A15Cs2}      \\
        & \VAFI1     &  11.783  &    818            &  0      &   (46)         & \onlinecite{A15Cs2}      \\
\hline
\end{tabular}
\end{center}
\label{tab_C60}
\end{table}

\begin{table}
\caption{Lattice parameters for pristine solid picene, coronene, and phenanthrene and superconducting transition temperature $T_c$ observed for doped systems.}
\vspace{0.1cm}
\begin{center}
\begin{tabular}{c@{\ \ \ \ \ }c@{\ \ \ }c@{\ \ \  }c@{\ \ \ }c@{\ \ \ }c@{\ \ \ }c}
\hline
              &   $a$  &  $b$  &  $c$  &  $\beta$     &  $T_c$  &\multirow{2}{*}{Ref.}  \\ 
              & (\AA)  & (\AA) & (\AA) &( $^{\circ}$ )&  (K)    &                       \\
\hline      
picene        &  8.480 & 6.154 &13.515 & 90.46        &  18,7   &\onlinecite{pic-st,pic-Kubozono} \\         
coronene      & 16.094 & 4.690 &10.049 & 110.79       &   15    &\onlinecite{cor-st,aroma-per} \\ 
phenanthrene  & 8.453  & 6.175 & 9.477 & 98.28        &   5-6   &\onlinecite{phe,phe2}    \\ 
\hline
\end{tabular}
\end{center}
\label{tab_aroma}
\end{table}

We performed DFT band calculation with {\em Tokyo Ab initio Program Package},\cite{TAPP} based on the pseudopotential plus plane-wave framework. We used the generalized-gradient approximation (GGA) exchange-correlation functional with the parameterization of Perdew-Burke-Ernzerhof\cite{PBE} and the Troulliar-Martins norm-conserving pseudopotentials\cite{TM} in the Kleinman-Bylander representation.\cite{KB} The pseudopotentials for alkali metals, K, Rb, and Cs were supplemented with partial core correction.\cite{PCC} The cutoff energies for wavefunctions and charge densities were set to 36 Ry and 144 Ry, respectively, and we employed 5$\times$5$\times$5 $k$-point sampling. We confirmed that this condition ensures well converged results.

The DFT calculations were performed for the following twelve materials: fcc K$_3$C$_{60}$, fcc Rb$_3$C$_{60}$, fcc Cs$_3$C$_{60}$ with three different lattice parameters, A15 Cs$_3$C$_{60}$ with four different lattice parameters, doped solid picene, coronene, and phenanthrene. The lattice parameters were taken from the experiments and internal coordinates were optimized.\cite{opt-note} In fcc A$_3$C$_{60}$, the disorder of the orientation of C$_{60}$ molecules was ignored, so the crystal symmetry is lowered from Fm$\bar{3}$m to Fm$\bar{3}$.

Before presenting the computational results, we summarize the basic properties of the compounds studied in the present paper. Table \ref{tab_C60} lists experimental values for the C$_{60}$ compounds, including the lattice constant $a$, the volume per \C60$^{3-}$ in solid,\cite{volume} applied pressure, and measured superconducting transition temperature $T_c$ or the N\'{e}el temperature $T_N$. The $a$ value and/or \C60$^{3-}$ volume can be controlled by the chemical and external pressures. In this table, the samples are arranged in the order of the increase of the lattice constant. For fcc \A3C60, $T_c$ first increases and reaches the maximum (35 K) around $a$=14.500 \AA. Then, it decreases down to $T_c\!\sim\!25$ K where the system experiences the metal-insulator transition (MIT) and becomes an antiferromagnetic insulator (AFI), for which the N\'{e}el temperature $T_N$ is around 2.2 K. 
A similar behavior is observed in the A15 system, while $T_N$ is significantly higher (46 K). This is because the A15 structure is bipartite and therefore less frustrated.\cite{Iwasa} Hereafter, we label the nine C$_{60}$ compounds as fcc-K, fcc-Rb, fcc-Cs(\Vopt1), fcc-Cs(\VMIT1), fcc-Cs(\VAFI1), A15-Cs(\Vsmall1), A15-Cs(\Vopt1), A15-Cs(\VMIT1), and A15-Cs(\VAFI1).

Table \ref{tab_aroma} shows the experimental lattice parameters for undoped solid picene, coronene, and phenanthrene and $T_c$ observed for doped systems. For doped solid picene, two different $T_c$ (18 K or 7 K) have been observed depending on the preparation conditions.\cite{pic-Kubozono} The superconductivity appears when the system is doped, but the details of the crystal structures in the superconducting phases have not been determined. Thus in the present study, the band calculations for aromatic compounds were performed for the artificially charged system where three negative charges per one hydrocarbon molecule were doped with a uniform compensating positive background charge. Hereafter, doped solid picene, coronene, and phenanthrene are referred to as solid \pic1, \cor1, and \phe1, respectively.

\subsection{Band structure and density of states}\label{sub_band}
\begin{figure*}[htbp]
\vspace{0cm}
\begin{center}
\includegraphics[width=1.0\textwidth]{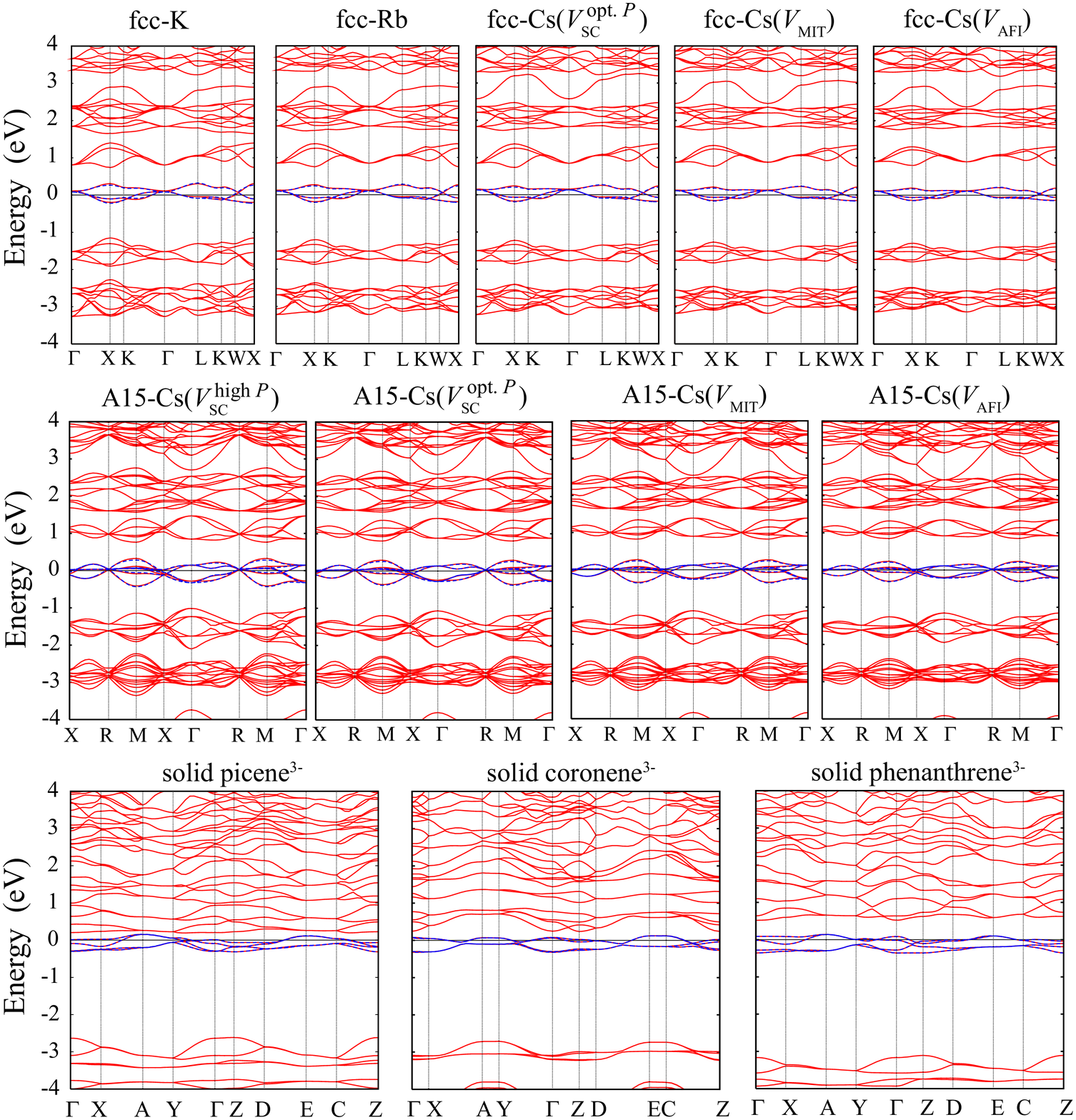}
\caption{(Color online) Calculated {\em ab intio} electronic band structure of fcc-K, fcc-Rb, fcc-Cs(\Vopt1), fcc-Cs(\VMIT1), fcc-Cs(\VAFI1), A15-Cs(\Vopt1), A15-Cs(\Vsmall1), A15-Cs(\VMIT1), A15-Cs(\VAFI1), solid \pic1, solid \cor1, and solid \phe1. In the case of aromatic compounds with monoclinic structure, 
the horizontal axis is labeled by the special points in the Brillouin zone with $\Gamma$, X, A, Y, Z, D, E, and C, respectively, corresponding to (0, 0, 0), (1/2, 0, 0), (1/2, 1/2, 0), (0, 1/2, 0), (0, 0, 1/2), (1/2, 0, 1/2), (1/2, 1/2, 1/2) and (1/2, 1/2, 1/2) in units of (${\bm a}^{\ast},{\bm b}^{\ast},{\bm c}^{\ast}$). The interpolated band dispersion with the derived tight binding Hamiltonian is depicted as blue dashed lines.    
 } 
\label{fig_band}
\end{center}
\end{figure*} 

Figure \ref{fig_band} shows our calculated GGA band structures for the fcc \A3C60 (upper 5 panels), A15 \Cs3C60 (middle 4 panels), and aromatic compounds (lower 3 panels). These compounds have common features in their band structure; i.e., we see narrow bands near the Fermi level separated from other bands, being preferable when we choose the target bands to construct an effective model. In the C$_{60}$ compounds, there are threefold degenerated states, which form the so-called `\t1u band' near the Fermi level, and we construct effective models for these bands. For aromatic compounds, the target bands are made from the lowest two unoccupied molecular orbitals (LUMO and LUMO+1) in an isolated molecule.~\cite{pic-Kosugi,phe-band,cor-band} 
It should be noted that unoccupied bands lie above the target bands more densely in the order of solid \pic1, solid \cor1, and solid \phe1. Since conduction bands can generate stronger screening when they reside closer to the target bands, we expect a weak repulsive interaction in solid picene$^{3-}$ compared to the other two.

\begin{figure}[htbp]
\vspace{0cm}
\begin{center}
\includegraphics[width=0.5\textwidth]{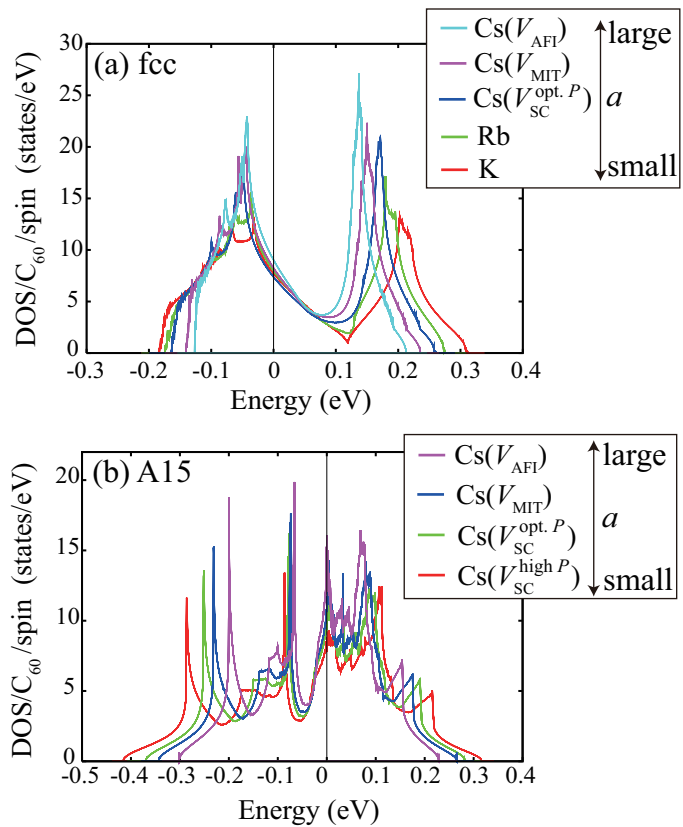}
\caption{(Color online) (a) Our calculated density of states (DOS) for \t1u band of fcc-K (red), fcc-Rb (green), fcc-Cs(\Vopt1) (blue), fcc-Cs(\VMIT1) (purple), and fcc-Cs(\VAFI1) (light blue). (b) DOS for \t1u band of A15-Cs(\Vsmall1) (red), A15-Cs(\Vopt1) (green), A15-Cs(\VMIT1) (blue), and A15-Cs(\VAFI1) (purple).}
\label{fig_dos}
\end{center}
\end{figure}

We show in Fig.~\ref{fig_dos} the calculated density of states (DOS) of the $t_{1u}$ band for fcc \A3C60 (a) and A15 \Cs3C60 (b). For both fcc or A15, the bandwidth $W$ monotonically increases as decreasing the lattice constant, but the DOS profile does not change drastically.  We list the values of $W$ in table~\ref{tab_W}. The bandwidth of A15 ($\sim$0.6 eV) tends to be larger than that of fcc ($\sim$0.4 eV), which is due to the difference in the inter-\C60 contact, i.e., `hexagon-to-hexagon' configuration for A15 and `bond-to-bond' one for fcc.~\cite{Cs3C60-band} It was found that the bandwidths of the aromatic compounds are nearly 0.5 eV (see table~\ref{tab_W}). 

\begin{table*}[htb] 
\ 
\caption{Calculated bandwidth $W$ of the target band and spatial Wannier spread $\Omega$ for twelve materials: fcc-K, fcc-Rb, fcc-Cs(\Vopt1), fcc-Cs(\VMIT1), fcc-Cs(\VAFI1), A15-Cs(\Vsmall1), A15-Cs(\Vopt1), A15-Cs(\VMIT1), A15-Cs(\VAFI1), solid \pic1, solid \cor1, and solid \phe1. For the aromatic compounds, the two value of $\Omega$ are listed; the left is the `lower-level' orbital and the right is `higher-level' one. Units are given in meV for $W$ and {\AA} for $\Omega$.}
\vspace{0.1cm} 
\centering 
\begin{tabular}{c ccccc c cccc c ccc}
\hline
&  \multicolumn{5}{c}{fcc \A3C60} && \multicolumn{4}{c}{A15 \Cs3C60} & 
& \multicolumn{3}{c}{aromatic compounds} \\ \cline{2-6}\cline{8-11}\cline{13-15}
& K & Rb & Cs(\Vopt1) & Cs(\VMIT1) &  Cs(\VAFI1)   &
& \Vsmall1 & \Vopt1 &  \VMIT1 & \VAFI1 && \pic1 &\cor1 &\phe1  \\ 
\hline
$W$  & \ \ 502\ \  &\  454\ \  &  427    &   379   &    341  &&    740  &   659   &   614  &   535   &&  477   &  447    &   505  \\ 
$\Omega$      &   4.28 &   4.21 &  4.19  &   4.14 &   4.10 &&   4.27 &  4.20  &  4.16 &  4.12 &&4.08, 4.13& 3.64, 3.67& 3.20, 3.08 \\ 
\hline
\end{tabular}
\label{tab_W}
\end{table*}

\subsection{Maximally localized Wannier orbitals}
\begin{figure}[htbp]
\vspace{0cm}
\begin{center}
\includegraphics[width=0.32\textwidth]{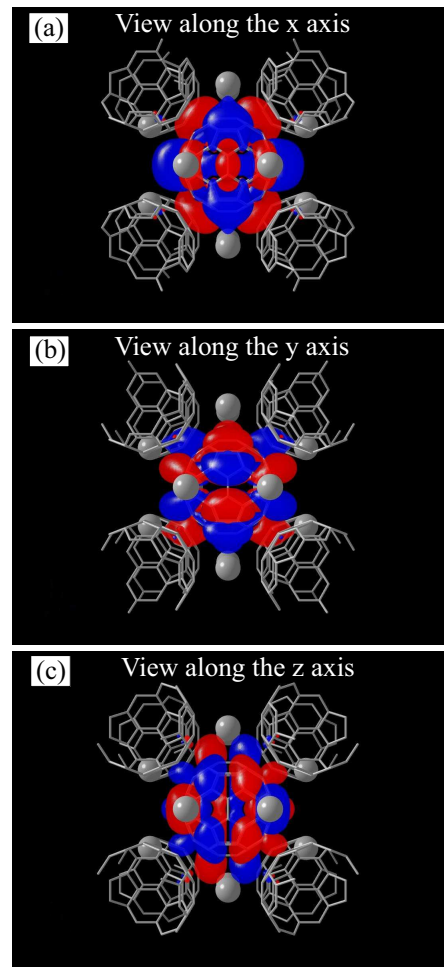}
\caption{(Color online) Isosurface of our calculated $p_x$-like maximally localized Wannier orbital of A15-Cs(\VAFI1) viewed along the (a) $x$ axis, (b) $y$ axis, (c) $z$ axes, drawn by VESTA.~\cite{VESTA} The red surfaces indicate positive isosurface and the blue surfaces indicate negative isosurface.}
\label{fig_w1}
\end{center}
\end{figure} 
\begin{figure}[htbp]
\vspace{0cm}
\begin{center}
\includegraphics[width=0.32\textwidth]{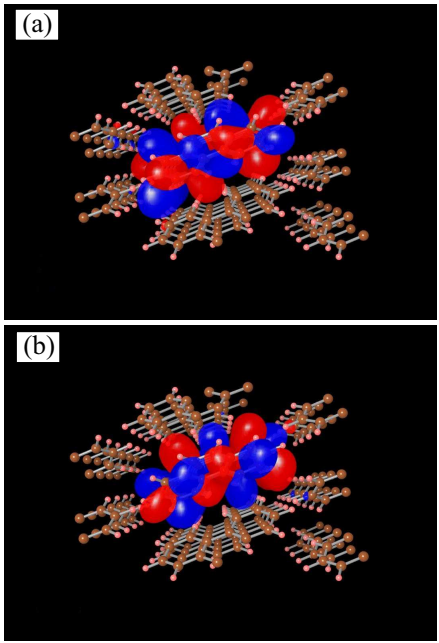}
\caption{(Color online) Isosurface of maximally localized Wannier orbitals of solid \phe1 with (a) lower onsite energy and (b) higher onsite energy, drawn by VESTA.~\cite{VESTA} The red surfaces indicate positive isosurface and the blue surfaces indicate negative isosurface.}
\label{fig_w2}
\end{center}
\end{figure} 
Figure \ref{fig_w1} shows a contour plot of one of MLWOs for the $t_{1u}$ bands of A15-Cs(\VAFI1). The results of other C$_{60}$ compounds are almost the same. From this figure we see that the resulting Wannier orbital is well localized at the single C$_{60}$ molecule. In this plot, we displayed the same orbital along the three directions; panels (a), (b), and (c) correspond to the view along the $x$, $y$, and $z$ axis, respectively. We see a node in the center of this orbital for (b) and (c), thus this orbital has $p_{x}$-like symmetry. Note that the view along the $y$ axis is not identical to the view along the $z$ axis, which is in contrast with the case of atomic $p$ orbitals. 
We note that the other two $p_{y}$- and $p_{z}$-like Wannier orbitals are symmetrically equivalent to the presented $p_{x}$-like orbital. We also note that the weight of the Wannier orbitals concentrates in the vicinity of the cage of the \C60 molecule and there is little weight inside it.

We next show in Fig.~\ref{fig_w2} a contour plot of two MLWOs of solid \phe1. In the aromatic compounds, the two basis orbitals of the effective model are not symmetrically equivalent and therefore we specify these orbitals as `lower' and `higher' orbitals in terms of the onsite level of MLWOs. The lower and higher orbitals are shown in the panels (a) and (b), respectively. We again see the resulting orbitals are well localized at the single molecules. The MLWOs of solid \pic1 and \cor1 are similar to those of undoped systems calculated in Refs.~\onlinecite{pic-Kosugi} and \onlinecite{cor-band}.

We list in table~\ref{tab_W} our calculated spatial spread $\Omega_{n}$ of MLWO for the twelve materials, where $\Omega_{n}$ is defined as
\begin{eqnarray}
  \Omega_{n} 
= \sqrt{\langle \phi_{n {\bf 0}} | r^2 | \phi_{n {\bf 0}} \rangle
- \bigl| \langle \phi_{n {\bf 0}} | {\bf r} | \phi_{n {\bf 0}} 
\rangle \bigr|^2}. 
\label{Spread}
\end{eqnarray}
In the \C60 compounds, the calculated Wannier spread is roughly 4 {\AA} and thus the estimated effective volume $\frac{4}{3}\pi\Omega^{3}$ is $\sim$268 {\AA}$^{3}$. The value is compared with the \C60$^{3-}$ volume listed in table~\ref{tab_C60} ($\sim$720-820 {\AA}$^{3}$), clearly indicating well-localized nature of MLWO on the single molecule.
We see that the Wannier spread has a weak positive correlation with the bandwidth $W$. In the aromatic compounds, the molecular size itself is different from each other, which result in the appreciable difference in $\Omega$. Note that, $\Omega$ has no clear correlation with $W$.  

\subsection{Transfer integrals}
Let us move on to transfer integrals. For the \C60 compounds, the band dispersion of the target band was found to be well reproduced only with nearest neighbor (NN) and next nearest neighbor (NNN) transfers.  The orbital index, 1, 2, and 3 denote $p_x$-, $p_y$-, and $p_z$-like orbitals, respectively. Onsite energies for three MLWOs are set to zero. From now on, `site' means one molecule and the coordinate of site ${\bf R}$ is defined as the center of the molecule. The transfer integrals $t_{nm}\!({\mathbf R})$ are represented as $3\times3$ matrix. In fcc (A15) structure, there are 12 (8) NN sites and 6 (6) NNN sites per site. From transfers to the specific site, other transfers to the equivalent sites are reproduced by proper symmetry operations. As a representative NN site, we choose ${\bf R}$=($R_x,R_y,R_z$)=(0.5, 0.5, 0.0) and (0.5, 0.5, 0.5) for fcc and A15 structure, respectively, where the coordinate is based on conventional cell. The transfer matrix to this site is 
\begin{eqnarray}
\label{eq-t1}
\left(
\begin{array}{ccc}
F_1 & F_2 & 0   \\
F_2 & F_3 & 0   \\
0   &  0  & F_4 \\
\end{array}
\right)
{\rm and} 
\left(
\begin{array}{ccc}
A_{1\ \ \ \ } & A_{2(3)\ } & A_{3(2)}  \\
A_{3(2)\ } & A_{1\ \ \ \ } & A_{2(3)}  \\
A_{2(3)\ } & A_{3(2)\ } & A_{1\ \ \ }  \\
\end{array}
\right)
\end{eqnarray}
for the fcc and A15 structure, respectively. 
In A15, the two \C60 molecules in the unit cell (denoted as A- and B-site) are not equivalent in terms of their orientations. So, in the matrix (\ref{eq-t1}), we show both the transfers from A-site to B-site and from B-site to A-site (in parentheses).
We choose ${\bf R}$=(1, 0, 0) for a representative NNN site for both structure, then the transfer matrix is written as 
\begin{eqnarray}
\label{eq-t2}
\left(
\begin{array}{ccc}
F_5 &  0  & 0   \\
 0  & F_6 & 0   \\
 0  &  0  & F_7 \\
\end{array}
\right)
{\rm and} 
\left(
\begin{array}{ccc}
A_{4\ } &  0  & 0   \\
 0  & A_{5(6)} & 0   \\
 0  &  0  & A_{6(5)} \\
\end{array}
\right)
\end{eqnarray}
for the fcc and A15 structure, respectively. 
The transfer matrix for A15 represents the A-A transfer and the B-B one (in parentheses).
Table \ref{tab_t1} shows the value of the parameters from $F_1$ to $F_7$ for fcc \C60 compounds. We note that $F_6$$\neq$$F_7$ is due to the lowering the symmetry from Fm$\bar{3}$m to Fm$\bar{3}$. Table \ref{tab_t2} shows the value of parameters from $A_1$ to $A_6$ for A15 \Cs3C60. For both systems, the values of hopping parameters decrease as the lattice parameters increases.

\begin{table}
\caption{Hopping parameters for fcc \A3C60 in Eqs.~(\ref{eq-t1}) and (\ref{eq-t2}). Units are given in $10^{-4}$eV.}
\label{tab_t1}
\begin{tabular}{c@{\ \ \ }r@{\ \ \ }r@{\ \ \ }r@{\ \ \ }r@{\ \ \ }r@{\ \ \ }r@{\ \ \ }r@{\ }}
\hline
                     & $F_1$ & $F_2$ & $F_3$ & $F_4$ & $F_5$ & $F_6$ & $F_7$  \\ 
\hline
fcc-K          & $-$40 & $-$339  &  421  & $-$187  & $-$94  &$-$14 & $-$2   \\ 
fcc-Rb         & $-$16 & $-$306  &  392  & $-$159  & $-$75  & $-$8  & 15   \\
fcc-Cs(\Vopt1) &  26   & $-$299  &  372  & $-$120  & $-$60  & $-$3  & 36   \\
fcc-Cs(\VMIT1) &  15   & $-$267  &  332  & $-$104  & $-$40  &  1   & 30   \\
fcc-Cs(\VAFI1) &  13   & $-$241  &  302  & $-$94   & $-$33  &  1   & 24   \\ 
\hline
\end{tabular}
\begin{center}
\end{center}
\end{table}
\begin{table}
\caption{Hopping parameters for A15 \Cs3C60 in Eqs.~(\ref{eq-t1}) and (\ref{eq-t2}). Units are given in $10^{-4}$eV.}
\label{tab_t2}
\begin{tabular}{c@{\ \ \ }r@{\ \ \ }r@{\ \ \ }r@{\ \ \ }r@{\ \ \ }r@{\ \ \ }r@{\ }}
\hline
                 & $A_1$ & $A_2$ & $A_3$ & $A_4$ & $A_5$ & $A_6$   \\ 
\hline
A15-Cs(\Vsmall1)     & $-$297  & 448   &  67   &  74   & $-$105 & $-$289      \\
A15-Cs(\Vopt1)       & $-$262  & 400   &  61   &  74   & $-$97  & $-$239     \\
A15-Cs(\VMIT1)       & $-$239  & 371   &  57   &  76   & $-$89  & $-$212      \\
A15-Cs(\VAFI1)       & $-$206  & 329   &  53   &  73   & $-$79  & $-$180     \\ 
\hline
\end{tabular}
\begin{center}
\end{center}
\end{table}

We next describe the procedure for obtaining transfers to the other NN sites or NNN sites. First we consider the NN case. For fcc structure, the transfer matrices to other five NN sites are written with $F_1$-$F_4$ as follows:
\begin{eqnarray*}
\left(
\begin{array}{rrr}
\phantom{ } & \phantom{FFF} & \phantom{FFF} \\[-4mm]
 F_4 &  0  & 0   \\
 0  & F_1 & F_2 \\
 0  & F_2 & F_3 \\
\end{array}
\right) \ 
&{\rm for}&\ {\bf R}=(0.0, 0.5, 0.5) 
\\
\left(
\begin{array}{rrr}
\phantom{ } & \phantom{FFF} & \phantom{FFF} \\[-4mm]
F_4 &   0  & 0   \\
 0  &  F_1 & -F_2   \\
 0  & -F_2 &  F_3 \\
\end{array}
\right) \  
&{\rm for}&\ {\bf R}=(0.0, 0.5, -0.5) 
\\
\left(
\begin{array}{rrr}
\phantom{ } & \phantom{FFF} & \phantom{FFF} \\[-4mm]
F_3 &  0  & F_2 \\
 0  & F_4 & 0   \\
F_2 &  0  & F_1 \\
\end{array}
\right) \ 
&{\rm for}&\ {\bf R}=(0.5, 0.0, 0.5) 
\\
\left(
\begin{array}{rrr}
\phantom{ } & \phantom{FFF} & \phantom{FFF} \\[-4mm]
 F_3 &  0  & -F_2 \\
  0  & F_4 &   0  \\
-F_2 &  0  &  F_1 \\
\end{array}
\right) \ 
&{\rm for}&\ {\bf R}=(-0.5, 0.0, 0.5) 
\\ 
\left(
\begin{array}{rrr}
\phantom{ } & \phantom{FFF} & \phantom{FFF} \\[-4mm]
 F_1 & -F_2 & 0   \\
-F_2 &  F_3 & 0   \\
  0  &  0   & F_4 \\
\end{array}
\right) \  
&{\rm for}&\ {\bf R}=(0.5, -0.5, 0.0).  
\end{eqnarray*}
For A15 structure, we have 
\begin{eqnarray*}
\left(
\begin{array}{rrr}
\phantom{F_{X(1)}} & \phantom{-F_{X(1)}} & \phantom{-F_{X(1)}} \\[-4mm]
 A_{1\ \ \ } &  A_{2(3)} & -A_{3(2)} \\
 A_{3(2)} &  A_{1\ \ \ } & -A_{2(3)} \\
-A_{2(3)} & -A_{3(2)} &  A_{1\ \ \ } \\
\end{array}
\right) \   
&{\rm for}&\ {\bf R}=(0.5, 0.5, -0.5) 
\\
\left(
\begin{array}{rrr}
\phantom{F_{X(1)}} & \phantom{-F_{X(1)}} & \phantom{-F_{X(1)}} \\[-4mm]
 A_{1\ \ \ } & -A_{2(3)} &  A_{3(2)} \\
-A_{3(2)} &  A_{1\ \ \ } & -A_{2(3)} \\
 A_{2(3)} & -A_{3(2)} &  A_{1\ \ \ } \\
\end{array}
\right) \ 
&{\rm for}&\ {\bf R}=(0.5, -0.5, 0.5) 
\\
\left(
\begin{array}{rrr}
\phantom{F_{X(1)}} & \phantom{-F_{X(1)}} & \phantom{-F_{X(1)}} \\ [-4mm]
 A_{1\ \ \ } & -A_{2(3)} & -A_{3(2)} \\
-A_{3(2)} &  A_{1\ \ \ } &  A_{2(3)} \\
-A_{2(3)} &  A_{3(2)} &  A_{1\ \ \ } \\
\end{array}
\right) \  
&{\rm for}&\ {\bf R}=(0.5, -0.5, -0.5).  
\end{eqnarray*}
The remaining transfers to the NN sites are reproduced by the relation $t_{nm}({\bf R})=t_{nm}(-{\bf R})$. 

Similarly, the transfers to the other NNN sites are described as
\begin{eqnarray*}
\left(
\begin{array}{ccc}
\phantom{AAA} & \phantom{A_X (A_{X(1)})} & \phantom{AAA} \\[-4mm]
F_7 (A_{6(5)}) &  0  &  0  \\
 0  & F_5 (A_4) &  0  \\
 0  &  0  & F_6 (A_{5(6)}) \\
\end{array}
\right) \  
&{\rm for}&\  {\bf R}=(0, 1, 0) 
\\
\left(
\begin{array}{ccc}
\phantom{AAA} & \phantom{A_X (A_{X(1)})} & \phantom{F_6(A_{5(6)})} \\[-4mm]
 F_6 (A_{5(6)}) &  0  &  0  \\
  0  & F_7 (A_{6(5)}) &  0  \\
  0  &  0  & F_5 (A_4) \\
\end{array}
\right)\  
&{\rm for}&\ {\bf R}=(0, 0, 1) 
\end{eqnarray*}
for fcc (A15) structure and the remaining NNN transfers are generated according to $t_{nm}({\bf R})=t_{nm}(-{\bf R})$. 

Using these NN and NNN transfer parameters, we construct the transfer part ${\cal H}_t$ in Eq.~(\ref{eq2}) of the effective model. The band dispersion for the \C60 compounds calculated from the resulting ${\cal H}_t$ is depicted as blue dashed lines in Fig. \ref{fig_band}, from which we see that the original GGA band dispersion is satisfactorily reproduced.

\begin{figure}[htbp]
\vspace{0cm}
\begin{center}
\includegraphics[width=0.45\textwidth]{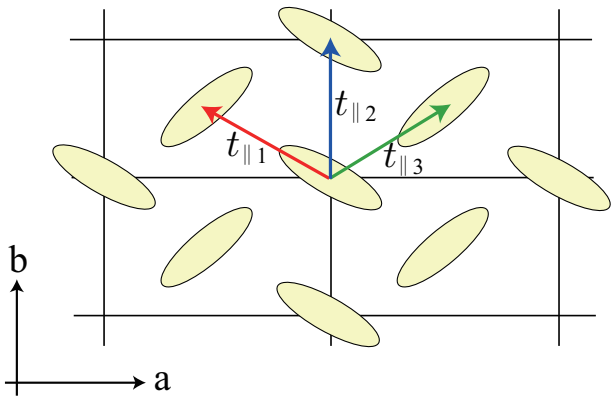}
\caption{(Color online) Schematic picture of intralayer transfer $t_{\parallel}$ along the three directions in aromatic compounds. The ellipses indicate molecules.}
\label{fig_tra}
\end{center}
\end{figure}

For the aromatic compounds, since there is no simple symmetry operation, their transfers are difficult to show concisely.\cite{pic-Kosugi,cor-band} For the aromatic compounds, we describe only some characteristic features of the transfers. The aromatic compounds are regarded as the stacking layered systems, so we expect 2-dimensional hopping structure. However, in the present transfer analysis, we found that the anisotropy of the transfers is not so simple. Table \ref{tab_t3} compares the maximum absolute value of the intralayer transfers ($t_{\parallel}$) with that of the interlayer transfers ($t_{\perp}$), where the intralayer is defined as the $ab$ plane. The intralayer transfers are further decomposed in the three directions and compared with each other (see Fig. \ref{fig_tra} for the definition of the three directions). 
The anisotropy ($t_{\perp}^{{\rm max}}/t_{\parallel}^{{\rm max}}$) is not so appreciable for solid \pic1, estimated as 20/59$\sim$0.34, and \phe1, as $\sim$0.49. In contrast, the anisotropy of \cor1 is significant and is $\sim$0.16. In the case of \cor1, the intralayer anisotropy is even strong $t_{\parallel 1}^{{\rm max}}/t_{\parallel 2}^{{\rm max}}$=7/87$\sim$0.08; this system is almost quasi-one-dimensional chain along the $b$ axis. We note that the original GGA band dispersion is well reproduced by short-range transfer hoppings (Fig. \ref{fig_band}).  
\begin{table}
\caption{Comparison between the maximum absolute values of intralayer transfer $t_{\parallel}$ and interlayer transfer $t_{\bot}$ for aromatic compounds. For the three directions in $t_{\parallel}$, see Fig.~\ref{fig_tra}. Units are given in meV.}
\label{tab_t3}
\begin{tabular}{c@{\ \ \ \ }r@{\ \ \ }r@{\ \ \ }r@{\ \ \ }r@{\ \ \ }}
\hline
            & $t_{_{\parallel 1}}^{\rm{max}}$ & $t_{_{\parallel 2}}^{\rm{max}}$ & $t_{_{\parallel 3}}^{\rm{max}}$  & $t_{\bot}^{\rm{max}}$ \\
\hline
solid \pic1 & 48 \ \  & 39 \ \ & 59 \ \ & 20 \ \  \\
solid \cor1 & 7  \ \  & 87 \ \ &  7 \ \ & 14 \ \ \\
solid \phe1 & 49 \ \  & 32 \ \ & 73 \ \ & 36 \ \   \\ 
\hline
\end{tabular}
\begin{center}
\end{center}
\end{table}

\subsection{Effective interaction parameters} \label{sub-int}
\begin{table*}[htb] 
\ 
\caption{$U$, $U^{\prime}$, $J$, and $V$ with three different screening levels [unscreened (bare), constrained RPA (cRPA), and fully-screened RPA (fRPA)] for the twelve compounds: fcc-K, fcc-Rb, fcc-Cs(\Vopt1), fcc-Cs(\VMIT1), fcc-Cs(\VAFI1), A15-Cs(\Vsmall1), A15-Cs(\Vopt1), A15-Cs(\VMIT1), A15-Cs(\VAFI1), solid \pic1, solid \cor1, and solid \phe1. For the aromatic compounds, the two value of $U$ are presented; the left is the `lower-level' orbital and the right is `higher-level' one. For `bare' and `cRPA' $U$, $U'$ and $V$ values, the unit is given in eV and $J$ is given by meV. For `fRPA', the unit is given in meV. In the bottom, we present our calculated cRPA macroscopic dielectric constant $\epsilon_{M}^{{\rm cRPA}}$ in Eq.~(\ref{epsilon}).}
\vspace{0.1cm} 
\centering 
{\scriptsize 
\begin{tabular}{c ccccc c cccc c ccc}
\hline
&  \multicolumn{5}{c}{fcc \A3C60} && \multicolumn{4}{c}{A15 \Cs3C60} & 
& \multicolumn{3}{c}{aromatic compounds} \\ \cline{2-6}\cline{8-11}\cline{13-15}
& K & Rb & Cs(\Vopt1) & Cs(\VMIT1) &  Cs(\VAFI1)   &
& \Vsmall1 & \Vopt1 &  \VMIT1 & \VAFI1 && \pic1 &\cor1 &\phe1  \\ 
\hline
$U_{\rm{bare}}$          & 3.27    &  3.31   &  3.32   &  3.35   &  3.37   &&   3.36   &  3.39   &  3.40   &  3.42   && 4.43,4.41 & 4.64,4.59 & 5.05,5.17  \\ 
$U^{\prime}_{\rm{bare}}$ & 3.08   &  3.11   &  3.12   &  3.15   &  3.17   &&    3.16   &  3.18   &  3.20   &  3.22   &&   3.55  &   4.33  &   4.55  \\ 
$J_{\rm{bare}}$          &   96    &  99     &   100   &  101    &  102    &&   97    &   99    &   100   &   101   &&   166   &   129   &   275   \\ 
$V_{\rm{bare}}$ &1.31-1.37&1.30-1.35&1.29-1.34&1.28-1.33&1.27-1.32&&1.37-1.38&  1.36-1.37&1.35-1.36&1.34-1.34&&2.08-2.32&2.79-2.84&2.29-2.43   \\ 
\hline
$U_{\rm{cRPA}}$          &  0.82    &  0.92    &  0.94    &  1.02   &  1.07   &&  0.93    &  1.02   &  1.07   &   1.14  && 0.73,0.74 &1.29,1.26&1.33,1.37   \\ 
$U^{\prime}_{\rm{cRPA}}$ &  0.76    &  0.85    &  0.87    &   0.94   &   1.00   &&  0.87    &  0.95    &   0.99   &   1.06  &&   0.58   &   1.15  &   1.17  \\ 
$J_{\rm{cRPA}}$          &   31    &   34    &  35     &   35    &   36    &&   30    &   36    &   36    &    37   &&   53    &    58   &   101  \\ 
$V_{\rm{cRPA}}$          & 0.24-0.25 & 0.26-0.27 & 0.27-0.28 & 0.28-0.29 & 0.30 && 0.30 & 0.31 &   0.32   &    0.34  && 0.26 & 0.59-0.60 & 0.47-0.48  \\ 
\hline
$U_{\rm{fRPA}}$          &   93    &  91     &  91     &   86    &    83   &&   107   &  102    &   99    &    93   && 155,151 & 149,120 & 166,172   \\ 
$U^{\prime}_{\rm{fRPA}}$ &   41    &   39    &  39     &   35    &    32   &&    50   &   45    &   42    &    37   &&   51    &   53    &   60   \\ 
$J_{\rm{fRPA}}$          &   25    &   26    &  26     &   26    &    25   &&    28   &   28    &   28    &    28   &&   38    &   39    &   57 \\ 
$V_{\rm{fRPA}}$          &   1-3   &   1-3   &  1-3    &   1-3   &    1-3  &&    2-3  &   2     &   2     &    1-2  &&   1-4   &   1-4   &    2  \\ 
\hline
$\epsilon_{M}^{{\rm cRPA}}$ &  5.6   &   5.1   &  4.9    &  4.6   &    4.4  &&  4.7  &   4.4   &   4.3   &   4.1  &&   12.0   &   5.5   &   6.3  \\
\hline
\end{tabular}
}  
\label{tab_e1}
\end{table*}

We performed RPA calculations to evaluate the screened Coulomb interaction $W({\bf r},{\bf r'})$ in Eqs.~(\ref{eq6}) and (\ref{eq7}), where the dielectric function was expanded in plane waves with an energy cutoff 7.5 Ry for fcc \A3C60 and aromatic compounds and 5.0 Ry for A15 \Cs3C60. The total number of bands considered in the polarization function was set to 335 (120 occupied+3 target+212 unoccupied) for fcc \A3C60, 670 (240 occupied+6 target+424 unoccupied) for A15 \Cs3C60, 310 (102 occupied+4 target+204 unoccupied) for solid \pic1, 315 (108 occupied+4 target+203 unoccupied) for solid \cor1, and 270 (66 occupied+4 target+200 unoccupied) for solid \phe1. 
The Brillouin-zone integral on wavevector was evaluated by generalized tetrahedron method.\cite{tetra} A problem due to the singularity of long-wavelength-limit Coulomb interaction in the evaluation of the Wannier matrix elements, $U_{nm}({\bf R})$ in Eq.~(\ref{eq6}) and $J_{nm}({\bf R})$ in Eq.~(\ref{eq7}), was treated in the manner described in Ref. \onlinecite{diele}.

The onsite interactions are specified by $U$=$U_{nn}({\bf 0})$ and $U'$=$U_{nm}({\bf 0})$ and $J$=$J_{nm}({\bf 0})$ for $n$$\neq$$m$. In the case of \C60 compounds, $U$, $U^{\prime}$, and $J$ take only one value according to the symmetry. For aromatic compounds, $U$ is different for two orbitals, so we present two values. We also denote the Coulomb repulsion between the neighboring sites as $V$. 

Table \ref{tab_e1} shows our calculated interaction parameters $U$, $U^{\prime}$, $J$, and $V$ with three screening levels (`bare', `cRPA', and `fRPA'). We see that the value of the Coulomb repulsion decreases as the screening processes increases. In the \C60 compounds, the bare value is $\sim$3.4 eV and after considering the screening by cRPA, the value is reduced to $\sim$1 eV. By taking account of the intra-target-band screening by fRPA, the value is further reduced to $\sim$0.1 eV. 
It should be noted here that the material dependence of the bare values in fcc or A15 is small; for example, 3.27 eV for fcc-K and 3.37 for fcc-Cs(\VAFI1). The difference is nearly 3 \%. This difference is ascribed to the difference in the spatial spread of MLWOs (see Table \ref{tab_W}). On the other hand, the material difference in the cRPA values is beyond 20 \%; 0.82 eV for fcc-K and 1.07 for fcc-Cs(\VAFI1). Indeed, this appreciable difference originates from the difference in the macroscopic dielectric constant defined as 
\begin{eqnarray}
  \epsilon_{M}^{{\rm cRPA}}
= \lim_{{\bf Q} \to 0} \lim_{\omega \to 0} 
 \frac{1}{ {\epsilon_{{\bf GG'}}^{{\rm cRPA}}}^{-1}({\bf q},\omega)  }
\label{epsilon}  
\end{eqnarray}
with $\omega$ being frequency and ${\bf Q}$=${\bf q}$+${\bf G}$, where ${\bf q}$ is a wave vector in the first Brillouin zone and ${\bf G}$ is a reciprocal lattice vector. We list the value in the bottom of the table. We see that the material dependence of $\epsilon_{M}^{{\rm cRPA}}$ is appreciable as 5.6 for fcc-K and 4.4 for fcc-Cs(\VAFI1), clearly indicating the importance of the screening effect in addition to the spatial Wannier spread.

For the aromatic compounds, differences in both the bare interaction and the screening effect contribute to the material dependence of the cRPA values; the bare interactions are
$U_{{\rm bare}}^{{\rm picene}^{3-}}$$\sim$
$U_{{\rm bare}}^{{\rm coronene}^{3-}}$$<$ 
$U_{{\rm bare}}^{{\rm phenanthrene}^{3-}}$
and after consideration of the cRPA screening, we obtain 
$U_{{\rm cRPA}}^{{\rm picene}^{3-}}$$<$ 
$U_{{\rm cRPA}}^{{\rm coronene}^{3-}}$$\sim$
$U_{{\rm cRPA}}^{{\rm phenanthrene}^{3-}}$.
Especially, in \pic1, the dielectric constant is markedly high as $\sim$12,\cite{eps_for_pic} making the cRPA $U$ value small appreciably.

We finally remark some points: As for the \C60 compounds, the equality $U^{\prime}$$\sim$$U$$-$$2J$ holds among effective parameters. This relationship also roughly holds for the aromatic compounds. The present $U$ value of cRPA for the \C60 compounds are small compared to the previous estimates of $U$ ($\sim$1-1.5eV).\cite{U1,U2,U3,U4} For all materials, 
the cRPA Coulomb interaction decays as $1/(\epsilon_{M}^{{\rm cRPA}}r)$ with $r$ being the distance between the centers of MLWOs, while the fRPA Coulomb interaction is limited to be short ranged due to the metallic screening (see table~\ref{tab_e1}). We note that the fRPA $U$ gives an opposite trend to the bare and cRPA $U$; for example, in the fcc \C60 compounds, the fRPA value slightly decreases as proceeding from fcc-K to fcc-Cs(\VAFI1). This is due to the fact that the Coulomb interaction is efficiently screened due to the increase in the density of states accompanied by the decrease of bandwidth. 
We also found that, in these systems, the exchange interaction $J$ are also efficiently screened; i.e., $J_{\rm cRPA}/J_{\rm bare}$$\sim$0.3. This makes a clear contrast to the case of the inorganic materials as $J_{\rm cRPA}/J_{\rm bare}$$\sim$0.8 such as $3d$ transition metals,\cite{cRPA-ex2} its oxides SrVO$_3$,\cite{cRPA-ex2} and iron-based superconductors.\cite{cRPA-ex6,cRPA-ex7}

\section{Discussions} \label{sec:discuss}
\subsection{Material dependence of effective parameters}

\begin{figure*}[t]
\vspace{0cm}
\begin{center}
\includegraphics[width=0.80\textwidth]{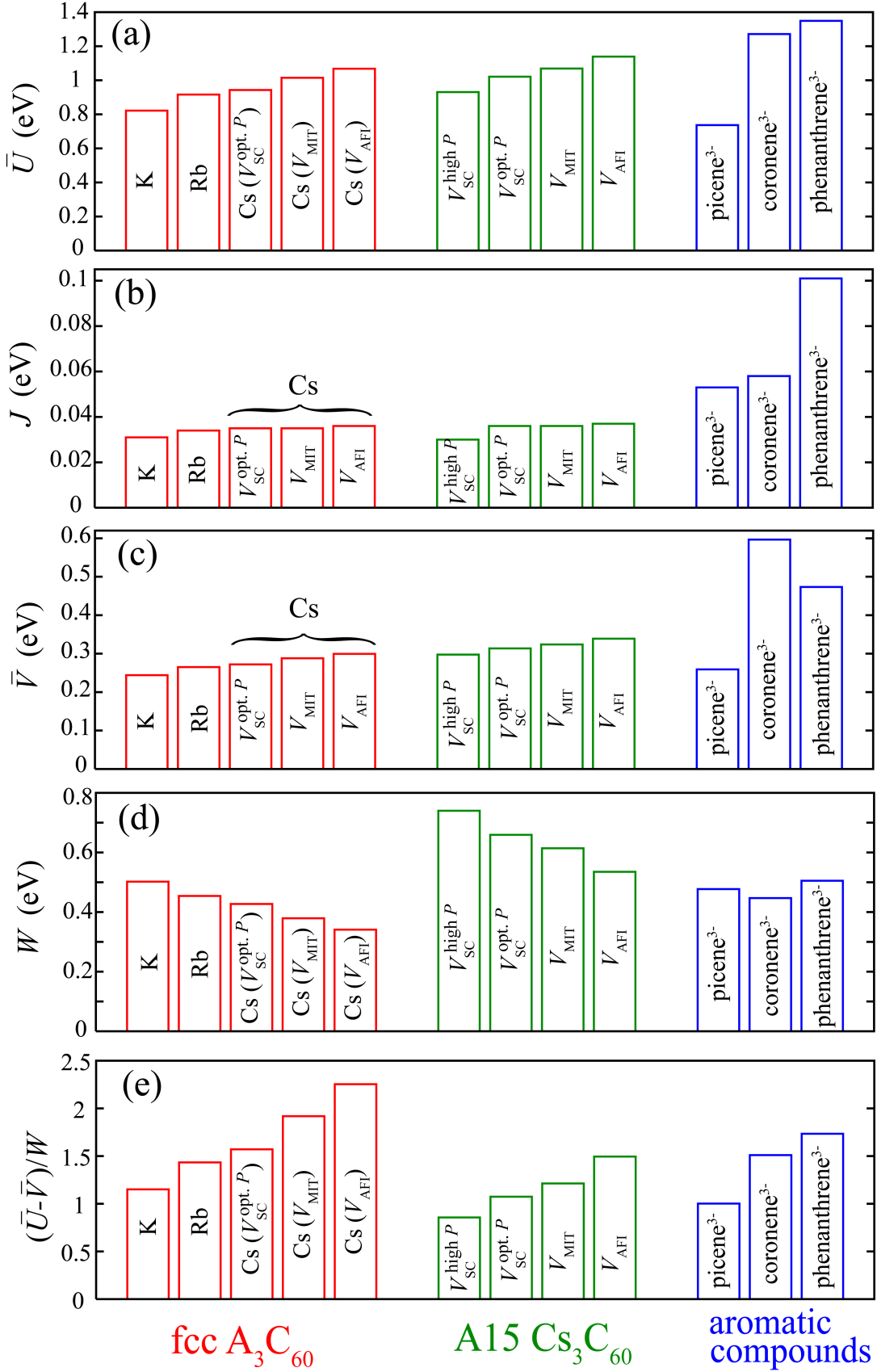}
\caption{(Color online) Material dependence of  
(a) the average of the onsite effective Coulomb repulsion $\bar{U}$, 
(b) the onsite effective exchange interaction $J$, 
(c) the average of the offsite effective Coulomb repulsion between neighboring sites $\bar{V}$, 
(d) the bandwidth of the target band $W$, and 
(e) the correlation strength $(\bar{U}-\bar{V})/W$.}
\label{fig_comp}
\end{center}
\end{figure*} 

Let us move on to the comparison of the effective interaction parameters among the twelve compounds. Figure \ref{fig_comp} summarizes the results of the cRPA calculation; the onsite Coulomb repulsion $\bar{U}$ averaged over MLWOs derived from the target band, the onsite exchange interaction $J$, the offsite interaction $\bar{V}$ averaged over nearest-neighbor sites, and the ratio $(\bar{U}-\bar{V})/W$ which measures the correlation strength in the system. Note that the net interaction is estimated as $\bar{U}-\bar{V}$, based on the analysis of the extended Hubbard model.

As for the \C60 systems, we see that $\bar{U}$ has appreciable material dependence, ranging from 0.8 eV to 1.1 eV [Fig. \ref{fig_comp}(a)]. This is ascribed to the differences in the size of Wannier orbitals and dielectric screening (see Sec. \ref{sub-int}). On the other hand, the material dependence of $J$ is weak and the value itself is negligibly small as $\sim$0.03 eV [Fig. \ref{fig_comp}(b)]. In general, small $J$ favors low-spin states, as is observed in experiments.~\cite{A15Cs2,fccCsN,Cs-NMR,A15-mag} It is interesting to note that there is a proposal that the Jahn-Teller coupling dominates over the Hund's rule coupling $J$, and induces superconductivity with the help of sufficiently large $U$.~\cite{Capone} Compared to $J$, we found that $\bar{V}$ is substantially large as $\sim$0.3 eV, being as large as $\sim$25\% of $\bar{U}$ [Fig. \ref{fig_comp}(c)]. 
As for $W$, which measures kinetic energy, we observed a decreasing trend as the lattice constant increases [Fig. \ref{fig_comp}(d)]. 
We also note that the energy scale for A15 \Cs3C60 is larger than that of fcc \A3C60 as mentioned in Sec.~\ref{sub_band}. The derived correlation strength of $(\bar{U}-\bar{V})/W$ exhibits a rather simple monotonic increasing behavior [Fig.~\ref{fig_comp}(e)], with the lattice constant increase. The presented $(\bar{U}-\bar{V})/W$$\sim$1 indicates that \C60 superconductors are categorized as strongly correlated electron systems.

For the aromatic superconductors, we found that the energy scale of $\bar{U}$ is similar to that of the \C60 superconductors [Fig. \ref{fig_comp}(a)]. On the other hand, it is interesting to note that the aromatic superconductors tend to have larger $J$ and $\bar{V}$ [Figs. \ref{fig_comp}(b) and (c)]. We can also see that the material dependence of the interaction parameters among the aromatic superconductors is also more significant, since the size and shape of the aromatic molecules are quite different from each other. As for $W$, they are similar for the aromatic and \C60 superconductors [Fig. \ref{fig_comp}(d)]. We found that aromatic superconductors are also in strongly correlated regime as \C60 ones, based on the analysis of the correlation strength 
[Fig.~\ref{fig_comp}(e)].

\subsection{Relation between electronic correlation and superconductivity}
\begin{figure*}[t]
\vspace{0cm}
\begin{center}
\includegraphics[width=1.00\textwidth]{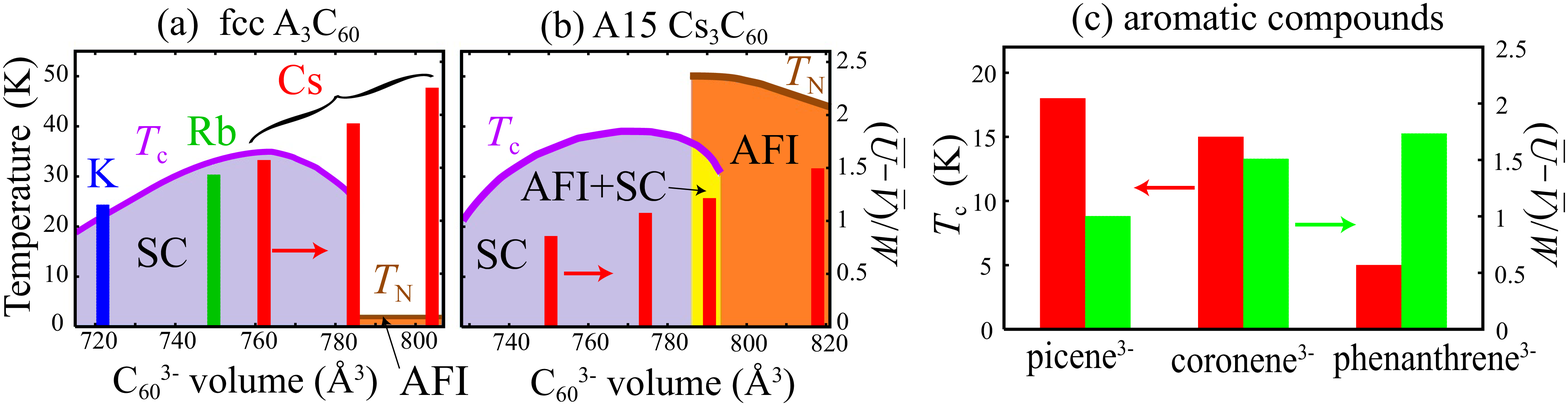}
\caption{(Color online) Relation between the experimental curve of superconducting or magnetic transition temperature ($T_c, T_N$) as a function of the C$_{60}^{3-}$ volume and the estimated correlation strength $(\bar{U}-\bar{V})/W$ (vertical bar): (a) fcc \A3C60 and (b) A15 \Cs3C60. For aromatic compounds (c), the measured $T_{c}$ in table~\ref{tab_aroma} and the calculated correlation strength are compared, where \pic1=(C$_{22}$H$_{14}$)$^{3-}$, \cor1=(C$_{24}$H$_{12}$)$^{3-}$, and \phe1=(C$_{14}$H$_{10}$)$^{3-}$. For the panels (a) and (b), the experimental phase diagram were taken from Ref.~\onlinecite{fccCsN} for fcc and Ref.~\onlinecite{A15Cs2} for A15. 
}
\label{fig_Tc}
\end{center}
\end{figure*} 

Next, let us discuss the relation between electronic correlation and superconductivity for the \C60 and aromatic superconductors. In Figs.~\ref{fig_Tc}~(a) and (b), we plot the superconducting transition temperature $T_c$ and the N\'{e}el temperature $T_N$ as a function of volume occupied per fulleride anion ($V_0$) for fcc \A3C60 and A15 \Cs3C60, respectively (see also table~\ref{tab_C60}). To see the relation between the electron correlation and the superconductivity we superpose a plot of $(\bar{U}-\bar{V})/W$ on the phase diagram. 
We see that while $(\bar{U}-\bar{V})/W$ and $T_c$ have a positive correlation up to $V_0$$\sim$760-770 \AA$^3$, for larger $V_0$, electron correlation becomes fatal for superconductivity and the system eventually becomes an insulator. We note that the critical value of $(\bar{U}-\bar{V})/W$ for the MIT sample is larger for fcc \A3C60 ($\sim$1.9) than A15 \Cs3C60 ($\sim$1.2). As discussed in Ref.~\onlinecite{Cs3C60-band}, it is important to consider the influence of lattice and orbital structure on MIT.~\cite{C60-MIT,Capone}

In Fig. \ref{fig_Tc}(c), we plot $T_c$ and $(\bar{U}-\bar{V})/W$ for the three aromatic superconductors, which shows a negative correlation. Therefore, it seems that electronic correlation does not favor superconductivity in these aromatic superconductors. Recently, doped 1,2:8,9-dibenzopentacene was found to have a quite high $T_c$$\sim$33 K. Since 1,2:8,9-dibenzopentacene is a bigger molecule than picene, coronene, and phenanthrene, 
the former interaction is expected to be small compared to the latter ones, reflecting the large Wannier spread of the 1,2:8,9-dibenzopentacene molecule. If there is no drastic change in the bandwidth $W$, which is probable in terms of the tendency shown in Fig. \ref{fig_comp}(d), the weakest electronic correlation will be realized in doped 1,2:8,9-dibenzopentacene. This trend is consistent with Fig. \ref{fig_Tc}(c). 

Regarding the role of electronic correlation in the \C60 and aromatic superconductors, there are two possibilities: The pairing mechanism in these compounds has a common root or these superconductors have completely different pairing glues. If we assume that the aromatic superconductors reside in the vicinity of the border between the superconducting and insulating phases, the first scenario is (at least partially) explicable to the behavior in Fig. \ref{fig_Tc}. 
On the other hand, in the second scenario, the electronic correlation enhances superconductivity for the C$_{60}$ compounds and inversely suppresses that for the aromatic compounds. In order to clarify this issue, experimental studies to determine the phase diagram for aromatic superconductors against temperature and volume occupied per anion are highly desired.\cite{note} Theoretically, microscopic calculations considering both electronic correlation and electron-lattice coupling are needed, which will be an interesting future problem.

\section{Summary} \label{sec:summary}
To provide insight into the role of electronic correlation in \C60 and aromatic superconductors, we derived effective models for wide range of the examples; fcc-K, fcc-Rb, fcc-Cs(\Vopt1), fcc-Cs(\VMIT1), fcc-Cs(\VAFI1), A15-Cs(\Vsmall1), A15-Cs(\Vopt1), A15-Cs(\VAFI1), solid \pic1, \cor1, and \phe1. To define the basis orbital of the effective model, we constructed MLWOs of isolated bands around the Fermi level. Transfer parameters are derived by evaluating the matrix elements of the Kohn-Sham Hamiltonian between the MLWOs. The low-energy electronic structures of the C$_{60}$ compounds are highly symmetric and isotropic, so that the original GGA band is reproduced with only 6 or 7 parameters. On the other hand, the aromatic compounds have quite anisotropic electronic structure. 

To quantify the strength of electronic correlation in these compounds, we estimated the effective interaction parameters 
such as $U$, $V$ and $J$, by means of the cRPA method. 
It was found that, in addition to the appreciable reduction of the diagonal part of the Coulomb interaction ($U$ and $V$), the off-diagonal part $J$ is also efficiently screened. Interestingly, 
all the \C60 and aromatic superconductors studied in the present work have a similar energy scale for the bandwidth and interaction parameters: $W$$\sim$0.5 eV, $U$$\sim$1 eV, $J$$\sim$0.05 eV, $V$$\sim$0.3 eV. This parameter range suggests that these compounds are a strongly correlated electron system.  
However, after examination of the material dependence, we found that a clear difference between the \C60 and aromatic compounds in the relation between electronic correlation strength and $T_c$; i.e., a positive correlation in the \C60 system and a negative correlation in the aromatic system. 

In the present study, we focused on the derivation for the electronic part of the effective model. For deep understanding of the low-energy physics for the carbon-based materials, however, the derivation of the electron-phonon interaction part is also imperative. The derivation for this part includes subtle problems on the definition of the basis for the phonon mode (Refs. \onlinecite{latticeWannier1} and \onlinecite{latticeWannier2}) and/or the exclusion of the double counting of the screening of the low-energy degree of freedoms, which needs the future studies.  

\begin{acknowledgements}
We thank Taichi Kosugi for providing us with the optimized structure data of undoped coronene and also for stimulating discussions. This work was supported by Grants-in-Aid for Scientific Research (No.~22740215, 22104010, 23110708, 23340095, 19051016), Funding Program for World-Leading Innovative R\&D on Science and Technology (FIRST program) on ``Quantum Science on Strong Correlation'', JST-PRESTO, 
the Strategic Programs for Innovative Research (SPIRE), MEXT, and the Computational Materials Science Initiative (CMSI), Japan.
\end{acknowledgements}


\begin{thebibliography}{99}
\bibitem{opt-note}
Initial internal coordinates for optimization are made after Ref.\onlinecite{pic-st} for solid picene and Ref.\onlinecite{phe-st} for solid phenenthrene.
\bibitem{volume}The volume per \C60$^{3-}$ is defined as $V_{\rm conv}/4$ for fcc structure and $V_{\rm unit}/2$ for A15 structure with $V_{\rm conv}$ ($V_{\rm unit}$) being the volume of conventinal (unit) cell.
\bibitem{eps_for_pic}
We note that the observed large $\epsilon_{M}^{{\rm cRPA}}$ of \pic1 ($\sim$12) originates from the polarization along the interlayer axis. The value of $\epsilon_{M}^{{\rm cRPA}}$ along the interlayer axis is 21.2 and those along the intralayer axis are 6.7 for the ${\bm a}^{\ast}$ axis and 8.2 for the ${\bm b}^{\ast}$ axis. The observed anisotropy is consistent with the result of Ref. \onlinecite{pic-pla}.
\bibitem{note}Indeed, the measurement of $T_c$ under pressure for alkali or alkali-earth metal doped phenanthrene was performed.\cite{phe,phe2}
\end{thebibliography}
\end{document}